\documentclass[11pt,twoside]{article}
  
\usepackage{asp2004}
\usepackage{graphicx}
\usepackage{epsf}
\usepackage{epsfig}
\usepackage{lscape}
\usepackage{amssymb}
  
\begin{document}
 
\title{Observational Review of this SINS Meeting} 
\author{Carl Heiles and Dan Stinebring} 
\affil{Astronomy Department, University of California, Berkeley CA;
  Physics \& Astronomy Deptartment, Oberlin College, Oberlin, OH}
  
\begin{abstract} 
We present a highly personal and biased review of the observational
portion of this meeting. We cover Tiny-Scale Atomic, Molecular, and
Ionized Structure (TSAS, TSMS, TSIS), emphasizing the physical aspects
involved when the structures are considered discrete.  TSAS includes
time- and angular-variable 21-cm and optical absorption lines.  TSMS
includes the Heithausen revolutionary component called Small Area
Molecular Structures. TSIS includes classical scintillation and discrete
structures that produce Arclets, Intra-Day and Intra-Hour Variables, and
Extreme Scattering Events. We conclude with reflections on the
relationship of Tiny-Scale Structure to the Local Bubble, TSAS to TSIS,
and the use of globular clusters as an illuminating backdrop.
\end{abstract} 
 
\section{Introduction}

We present here a highly biased observational review of this
meeting. Rather than giving an equally-weighted recap, we emphasize
those aspects that happen to be of most interest---to us. In the
process, we can't help slighting some of the interesting and important
papers presented at the meeting.  In particular, we are more interested
in tiny-scale structures as discrete structures instead of part of a
statistical distribution of turbulence. Our presentation, and the detail
and accuracy of our review, reflects our biases.

While the presence of tiny-scale structure is empirically fascinating,
that's not why this subject and this meeting are important. Rather, for
tiny, discrete structures to manifest themselves observationally, they
have to be {\it dense}; otherwise you can't build up the high observable
column densities in short path lengths. And these high densities mean
{\it high pressures}. This is a problem because these highly
overpressured structures want to explode!

There are two types of overpressures, one with respect to the local
environment and one with respect to the overall Galactic hydrostatic
equilibrium. Locally, typical thermal pressures in the ISM (which we
always write as $\widetilde {P} \equiv {P \over k}$, with units of
cm$^{-3}$ K) are a few thousand. Turbulence, magnetism, and cosmic rays
all have comparable pressures, so the total is several times the thermal
pressure. In the final analysis, the ensemble/time average of the total
pressure has to be the hydrostatic-equilibrium pressure for the vertical
structure of the Galaxy, which is $\widetilde{P} \sim 28000$ cm$^{-3}$ K
(Boulares \& Cox 1990). The inferred pressures of tiny-scale structures
as discrete objects dwarf this equilibrium pressure, so they cannot
possibly be confined for long. This pressure problem lies at the heart
of the difficulties in understanding tiny-scale structure.

The above paragraph assumes that the observational manifestation of
tiny-scale structure---variability---comes from discrete
structures. Another possibility is that the variability arises simply
from the accumulated small-scale turbulence along the line of sight
(Deshpande, SINS). Another is that velocity gradients plus scintillation
lead to apparent variations in optical depth (Gwinn, SINS). In other
words, the observed effects might result from actual structures
(``things'') or, more elegantly, from ordinary turbulence piled up along
the line of sight (``spooks''). These appropriately descriptive terms were
introduced by Vladimer Strelnitski (SINS), who discussed the ``things''
called interstellar masers.

A convention: we refer to lots of papers presented at the meeting, which
we identify by placing (SINS) after the speaker. Personal comments about
neutral gas are Carl's, while those about ionized gas are Dan's.

\section{ Tiny-Scale Atomic Structure (TSAS)}

\subsection{ The Ubiquity of TSAS: Historical Perspective}

\subsubsection{ 1997}

Tiny-Scale Atomic Structure (TSAS) is revealed by variability of 21-cm
and optical absorption lines with both position and time. The story
began with the original VLBI effort by Dieter, Welch, \& Romney
(1976). Given our knowledge of the interstellar medium, it seemed
impossible that there could be observable tiny-scale structure because
it would mean high volume densities, and therefore pressures. Privately,
I scoffed at the possibility. But fortunately, the spirit of adventure
prevailed and here we are today as a result. The TSAS had inferred sizes
of $\sim 30$ AU; the pressure problem was real, but could be ameliorated
by low temperatures and anisotropy (Heiles, SINS).

In 1997 (review by Heiles 1997; H97) there were VLBI studies of three
sources, 3C138, 3C147, and 3C380; the first two, and probably 3C380,
showed TSAS. There were searches for time variability of the 21-cm line
absorption against 6 pulsars; all showed variability. There were
searches for angular changes in optical absorption lines against 17
binary stars with separations ranging down to 450 AU; all showed
changes. 

TSAS was {\it ubiquitous}. Adding spice was Tiny-Scale Ionized Structure
(TSIS), which was revealed by Extreme Scattering Events (ESEs), and
Tiny-Scale Molecular Structure (TSMS), which was revealed by time variability
of H$_2$CO absorption lines. TSIS and TSMS might not have been
ubiquitous, but they existed.

\subsubsection{ 2006}

How things have evolved! TSMS and TSIS have flourished, with the
discovery of fascinating new effects and structures (\S \ref{tsms} and
\S \ref{tsisobs}). In TSAS, more pulsars have been observed for 21-cm line
variability and some of the original variable ones have been
reobserved. The result: most pulsars do {\it not} exhibit time-variable
21-cm line absorption (Weisberg \& Stanimirovic, Minter \& Balser,
SINS). Seven sources have been studied with VLBI (Brogan, SINS; Brogan
et al.\ 2005) and only two exhibit TSAS. 3C138 stands out very
distinctly as the best case---it exhibits variability in {\it both}
angle and time! VLBI maps are compelling in revealing it to be a single
physical structure (a ``thing''); 3C147 shows detectable structure that
might be from turbulence (a ``spook'') instead of a single
structure. See Brogan (SINS) for details and references.

So now, in 2006, TSAS, as revealed by the 21-cm line, is {\it by no
means} ubiquitous!  Optically, though, structure remains ubiquitous. And
TSIS is alive and thriving. TSMS is not only still there, but is also
both more puzzling than ever and also more indicative of physical
processes in turbulence (\S \ref{tsms}).

TSAS has sizes measured in tens of Astronomical Units (AU). We also find
very small structure that isn't quite so tiny, as reviewed by
Stanimirovic (SINS). She discusses very low-column density clouds whose
inferred sizes are $\sim 2000$ AU, lots bigger than the TSAS but,
nevertheless, small in anybody's book. There's a large range of
unstudied scale sizes lying between ``tiny'' and typical parsec-scale
structures.

\subsection{ TSAS Optical Lines}
\vspace{0.08in}

Lauroesch (SINS) reviewed the observations of NaI absorption lines
against stars. These lines have it all: time variability, angular
variability from two-point sampling of binary stars, and angular
variability from quite densely sampled stars in clusters.

\subsubsection{ Angular Variability of Optical Lines}

Angular differences of NaI absorption against binaries are {\it always}
seen (Watson \& Meyer 1996). The projected binary separations run from
500 to 30000 AU.  Against the globular cluster M14 the sampling is dense
enough to make a good map; variations by a factor of 16 occur on scales
of 4 arcsec (Meyer \& Lauroesch 1999). The absorbing material lies in
the Galactic plane, not in the cluster itself, so it's distance is
probably $\sim 100$ pc, which translates to a variability length scale
of $\sim 500$ AU. Well-sampled absorption line changes are also seen
against the Galactic clusters h and $\chi$ Persei (Points, Meyer, \&
Lauroesch 2004).

\subsubsection{ Time Variability of Optical Lines}

Time variability is definitely {\it not} ubiquitous, with only 10
detections out of at least 40 stars examined. Some of these changes are
really dramatic: for example, the absorption against the star 23 Ori,
which itself moves only 0.8 AU per year, {\it doubles every year!}.

Crawford (2003) presents an excellent comprehensive review of both the
data and their interpretation. All but one of these time-variable
absorption line stars lie behind shell structures, seen either as
supernovae remnants or as expanding HI bubbles in the 21-cm line, so
it's not just the motion of the star but also the motion of the
foreground gas itself that produces rapid variability. The one exception
is $\kappa$ Vel; despite its location in the constellation Velorum, it
lies well away from both the Vela supernova remnant and the Gum nebula,
so the variablity arises from the star's motion alone.

\subsubsection{ Are Optical Variations from Column Density or Ionization?}

        Common optical interstellar absorption lines are produced by
minority ionization species.  NaI is the most convenient, and
consequently its line is the one that usually reveals time- and angle-
variability. As a minority species, its line strength $\propto
T^{-1.6}$ (Heiles, SINS). TSAS should be cold and dense, so these lines
should be strong. Moreover, the line strength is sensitive to
temperature and density, so its variations with angle and time might be
tracing small changes in environmental conditions instead of large
changes in total column density of Na (and H) nuclei.

There are two arguments that NaI variations are primarily tracing
relatively small changes in environmental conditions: \begin{enumerate}

\item Majority ionization species are better tracers of total column
  density than minority ones. Majority ionization species are less well
  studied, but when they are observed they rarely exhibit variations
  with angle or time; for example, see Lauroesch et al.\ (1998).

\item The CI fine structure lines are very good tracers of interstellar
  pressure (Jenkins \& Tripp 2001; Jenkins, SINS).  Lauroesch et al.\
  (2000) and Welty \& Fitzptrick (2001) observed CI lines that
  correspond to time-variable NaI lines, but find no unusually large
  pressures.

\end{enumerate} 

\noindent Both arguments favor the idea that NaI variability results
from small changes in environmental conditions instead of large changes
in column density. 

However, the jury is still out! Crawford (2003) explicitly discusses the
important caveat: multiple regions along the line of sight tend to mask
high-pressure low-column density regions, and some of the observed lines
do indeed look like they have multiple components. Jenkins (SINS)
interprets his recent data as supporting the presence of high-pressure
regions (\S \ref{jenkins}); and Crawford argues quite
convincingly that two stars (HD32040 and HD219188) do, indeed, indicate
high-pressure regions.

It seems that the most straightforward conclusion is that most of the optical
variability comes from ionization effects---but also that the most
straightforward conclusion may not be correct, at least in some
cases. We clearly need lots more work in this area!

\subsection{ The Bimodal CNM Pressure Distribution} \label{jenkins}

This is very relevant to TSAS because of the TSAS pressure problem.
Jenkins (SINS) follows the techniques of Jenkins \& Tripp (2001)
to derive pressures for about 100 stars. Their technique exploits the
relative populations of the three fine structure spectral lines
associated with the $^3$P CI ground state.  The populations of the three
levels depend on collision rates, and comparing the three line
intensities provides information on density and temperature in a form
that approximates pressure. Thus CI is a probe of interstellar pressure.

Jenkins finds that the probability density function of the pressure is
bimodal. The primary component, which contains most of the mass, is
centered near $\widetilde{P} \sim 3000$ cm$^{-3}$ K. This is undoubtedly
the standard CNM, whose statistical properties are given by Heiles \&
Troland (2005). The other is centered near the amazingly high
$\widetilde{P} \sim 10^6$ cm$^{-3}$ K.

This high-pressure component is just what we need for TSAS!  Instead of
invoking the physically-reasonable pressure equality between TSAS and
other interstellar components, together with the necessary adoption of
unusual temperatures and geometries, we might simply throw caution to
the winds and accept the high pressure! For these $\widetilde{P} \sim
10^6$ cm$^{-3}$ K components, Jenkins suggests that the temperature must
exceed 100 K. This greatly increases (by a factor $\sim 50$) the
quasi-equilibrium TSAS column density at $T \sim 16$ K.  If indeed there
are components with this pressure and temperature, then Heiles's (SINS)
equation (6) implies the components are roughly spherical!

So we have two contrasting ideas. \begin{enumerate}

\item The H97 picture of approximate
{\it pressure equality}, minimizing the TSAS pressure by a invoking a
combination of the lowest possible TSAS temperature (16 K) and
morphological {\it anisotropy}. 

\item In contrast, we have the Jenkins picture of huge {\it
overpressure} and much warmer TSAS ($\gtrsim 100$ K), and approximate
morphological {\it isotropy}. How different can you get?
\end{enumerate}

The Jenkins picture produces a severe overpressure for the TSAS, and any
such structure would explode at Mach 20 or so. It couldn't last long and
the ensemble of observed structures would be continually dissolving and
reforming. So from the standpoint of {\it theoretical interpretation},
it's hard to envision. However, from the standpoint of {\it pure
empiricism}, this isn't a problem because both the high pressures and
the TSAS time variablity are {\it observed}---and correct observations
always trump theory.  

Nevertheless, I am bothered: from the conceptual framework of
astrophysics, it's hard for me---even as an {\it observer} by trade---to
imagine a mechanism by which such structures can actually be produced at
a sufficient rate that we see them reasonably often, and I keep looking
for ways to discount Jenkins's results. This reminds me of the situation
I described in the first paragraph of this paper: I'm privately scoffing
at the possibility that Jenkins's high pressures are correct. And, of
course, I was wrong then\dots

\subsection{ Extinction against Globular Clusters---Hot Off the Press!}

As a side project, Ivan King (private communication) is using the
turnover region of the HR diagram, whose tracks are roughly orthogonal
to the reddening vector, to derive accurate reddenings of individual
stars. This works well for mapping the reddening against globular
clusters, which contain lots of evolved stars. He has already made maps
for about 20 clusters, and intends to double that number!

He typically sees reddening structure at the level of a few hundredths
of a magnitude in E(B-V), or $N(HI) \sim 6 \times 10^{19}$ cm$^{-2}$, on
20 arcsec scales (but so far, at least, not below). If the absorbing
material is 100 pc distant, this corresponds to a few thousand AU. These
are extinctions, so they reflect {\it total} column densities.  If this
variability is produced by isotropic structures, then the volume density
$n(HI)$ is equal to the column density divided by the length scale,
which is $n(HI) \sim 4000$ cm$^{-3}$; at any reasonable interstellar
temperature, this produces a hefty overpressure!

It would be very interesting to compare the NaI maps with the
total extinction maps; this would provide a definitive measurement of
the degree to which the NaI variations are produced by ionization
structure. Here at this meeting we are focused on tiny-scale structure
in {\it total} density, but clearly tiny-scale structure in {\it
ionization fraction} opens up a new window on the ISM. And comparison of
these two datasets provides both!

\subsection{ A True Power Law for CNM Structure?}

Optical images of reflection nebulae have the capability of revealing
small-column density structures. A spectacular example is the Pleiades,
where a network of very fine-scale aligned filaments is seen in
scattered light. The alignment suggests a magnetic field.  It is thought
that the cluster and the ISM encountered each other by chance and that
the ISM has suffered only minimal interaction with the cluster, so we
have a nice unbiased sample of tiny-scale structures in the ISM that
happen to lie ``under a streetlight''.

Gibson (SINS) combined optical and radio images of the Pleiades optical
reflection nebulosity over resolution scales ranging from $\sim 0.1$
arcsec (HST PC image), to $\sim 1$ arcsec (WIYN telescope image with
excellent seeing), to $\sim 10^4$ arcsec (a mosaic of Burrell Schmidt
images), to somewhat larger (VLA D-array image with zero-spacing from
the GBT). Combining all these into a single power spectrum yields a
power-law index $-2.8 \pm 0.1$---this is over 5 orders of magnitude in
scale!

This is close to Kolmogoroff, like the famous Armstrong, Rickett, \&
Spangler (1995) spectrum, which covers even more range but is patched
together from disparate data sources and astronomical objects. It is
also the same slope found by Deshpande, Dwarakanath, \& Goss (2000) in
their VLA map of Cas A. The similarity of all these slopes suggests that
this slope is a robust characteristic of the ISM. Clearly, such studies
of more regions would confirm (or deny) this suggestion.

But suppose, after more studies, that the slope turns out to be about
the same everywhere. Then we have to ask: so what? If the same slope
applies to disparate types of region, does it provide any useful
physical information or physical insight?

\subsection{ The Reigel-Crutcher ``Cloud'': A Causal Structure}

McClure-Griffiths (SINS) showed a fabulous map of the Riegel-Crutcher
cloud, which is a cold cloud seen in self-absorption discovered decades
ago. Her new map uses the SGPS data to map a huge section of the cloud
with exquisite angular resolution ($\sim 100$ arcsec). The images show
angel-hair structure with clear filaments whose widths are unresolved at
about 0.07 pc resolution. With column densities $\sim 10^{20}$
cm$^{-2}$, the volume density $\sim 400$ cm$^{-3}$ if they are
cylindrical. With the inferred temperature $\sim 40$ K, this leads to
overpressures with $\widetilde P \sim 16000$ cm$^{-3}$ K.

These overpressures are not unexpected because this cloud is associated
with the edge of the North Polar Spur, which is an expanding
superbubble. From X-ray data (Nousek et al.\ 1982), one can estimate the
internal hot-gas pressure to be $\widetilde{P} \sim 5 \times 10^4$
cm$^{-3}$ K. This hot, overpressured gas drives the shock, which sweeps
up the gas into the cold shell containing this cloud. The cloud pressure
should be comparable to the hot gas, but it's lower.

This isn't so bad, though, because the filamentary angel-hair structure
suggests a magnetic field, which adds pressure. McClure-Griffiths uses
optical polarization of background stars with the Chandrasekhar-Fermi
method to infer a magnetic field strength of $\sim 40$ $\mu$G in the
filaments; this is large, far larger than the hot gas pressure! Pressure
equality would want a field strength $\sim 10$
$\mu$G. Chandrasekhar-Fermi field strengths are not very accurate, so
maybe it all works out. In any case, any overpressure problem here is
small compared to the pressure problems posed by the discrete-cloud
interpretation of the TSAS.

This is a cold cloud whose origin is clear: it's the shell
associated with the expanding North Polar Spur superbubble, and the
cloud gas was swept up by the superbubble shock. This cloud is nearby
and presents an ideal opportunity to study the detailed structure of a
supershell wall. In particular, we often think that TSAS (and TSIS)
might be associated with shocks, and this is an ideal opportunity to
find out.  With its well-defined self-absorption 21-cm line and
large suggested magnetic field, it's an ideal candidate for Zeeman
splitting.

\subsection{ Summary: Four Competitors for TSAS}
\vspace{0.1in}

We have four different ideas for TSAS: \begin{enumerate}

\item TSAS is discrete structures in approximate thermal pressure
  equality with other ISM phases (Heiles, SINS). To achieve this, the
  TSAS temperatures must be cold, approaching the lowest possible limit
  of 16 K. And it requires nonspherical structures with morphological
  anisotropy factor $G\sim 7$ or so.

\item Velocity gradients plus scintillation lead to apparent variations
in optical depth of small HI structures (Gwinn, SINS).

\item TSAS is simply a result of the pileup of column density versus
  velocity along the line of sight that results from the turbulent
  cascade spectrum (Deshpande, SINS). The spectrum needs to be flat
  enough to provide enough small-scale structure. There are probably
  some lines of sight where this requirement is satisfied.

\item TSAS can result from discrete objects or turbulent fluctuations
  that are highly overpressured (Jenkins, SINS). Temperatures are
  $\gtrsim 100$ K, somewhat warmer than standard CNM temperatures. If we
  have discrete structures, they can be round.

\end{enumerate}

It wouldn't be surprising for all ideas to apply, with one idea
dominating a particular sightline depending on conditions.  But\dots how
to empirically decide in any particular case?

\begin{enumerate}
\item If discrete objects have sharp edges, they can be recognized by
  mapping the TSAS absorption against background extended sources. In
  the optical we can do this with star clusters and in the radio with
  VLBI against background radio sources. VLB maps of 3C138 show a
  spectacular discrete object. The concept of TSAS being discrete objects
  is the simplistic, na\"ive observer's view, which is why it appeals to
  me. 

\item If TSAS simply results from interstellar turbulence, then we
  expect the slope of the turbulent spectrum to be flatter in regions
  that exhibit TSAS. We no longer regard TSAS as being ubiquitous, so if
  the turbulent slope is the same everywhere---as might be suggested by
  Gibson's (SINS) results---then this would argue against TSAS being
  simply turbulence.

\item If TSAS is highly overpressured as suggested by Jenkins (SINS),
  then the CI line ratios should reflect this so there should be an
  observable correlation between high pressures and the presence of
  TSAS. Existing studies don't show such a correlation, but we must
  emphasize the caveats that Crawford (2003) has suggested.

\end{enumerate}

\section{Tiny-Scale Molecular Structure (TSMS) \label{tsms}}

\subsection{TSMS in Ordinary Molecular Clouds}

Marscher, Moore, \& Bania (1993) and Moore \& Marscher (1995) detected
TSMS in ordinary molecular clouds by time variability of the 6-cm
H$_2$CO lines seen in absorption against quasars. The inferred
transverse sizes $\sim 10$ AU and densities $n(H) \gtrsim 10^6$
cm$^{-3}$. With $T \sim 10$ K these densities give $\widetilde P \gtrsim
10^7$ cm$^{-3}$ K, which greatly exceeds standard pressures in the
ISM. However, this huge overpressure does not necessarily imply explosive
expansion because self-gravity is an important force in molecular
clouds.

\subsection{ Small Area Molecular Structures (SAMS)---a 
Spectacular New Form of Molecular Cloud}

Heithausen (SINS) reviewed his remarkable Small Area Molecular
Structures (SAMS). These are small, dense Galactic molecular clouds,
which he discovered serendipitously while mapping mm-wave molecular
lines in tidal features of the M81 group (Heithausen 2002). With overall
sizes $\sim 1$ arcmin, we cannot classify an individual SAMS as ``tiny''
in the sense used for TSAS. However, the SAMS contain substantial
substructure (Heithausen 2004) at the few arcsec scale, equivalent to a
few hundred AU at a distance of 100 pc: that certainly classifies them
as tiny! 

An unbiased search (Heithausen 2006) shows that they might be very
common.  Dirsch, Richtler, \& G\'omez (2005) serendipitously found one
optically in projection against NGC3269, which shows that column
densities are high.  So far, these objects have been seen only at high
Galactic latitudes, where their weak emission doesn't have to compete
with the much stronger CO emission at lower latitudes. They are not
embedded in dense HI and are best described as lying ``in the middle of
nowhere''; it would be like having a city block of New York City in the
middle of New Mexico.

With volume densities of H-nuclei $n(H) \sim$ several thousand
cm$^{-3}$, these objects are heavy and should fall towards $z=0$, yet
their velocities are low ($\lesssim 10$ km s$^{-1}$) and they seem to be
plentiful. Maybe they are held together and/or are prevented from
falling by a magnetic field, as suggested by Don Cox.  What makes them?
What's their life cycle like? Why do their interiors have such small
substructure at the TSAS size scale? As such, we can call them TSMS, and
in some respects they are even more challenging than TSAS.

\subsection{TSMS and Dissipative Turbulence}

Recent studies of TSMS by the French group concentrate on macroscopic
velocity gradients and their associated turbulent dissipation on
nonequilibrium chemistry. These studies are detailed by Hily-Blant
(SINS) and Falgarone (SINS). They used the IRAM 30 m telescope to map
$^{12}$CO and $^{13}$CO in the Taurus molecular clouds and found
filaments with diameters ranging down to $\lesssim 1000$ AU, which size
is limited by the angular resolution. These filaments follow the local
magnetic field orientation as derived from starlight polarization.  The
filaments have H$_2$ volume densities ranging up to $\sim 2000$
cm$^{-3}$, with $T \sim 8$ K. These conditions are not particularly
extreme for molecular clouds and do not produce a severe overpressure.

What is very unusual, however, is the macroscopic velocity field
associated with the filaments.  The filaments exhibit intense velocity
shear in the form of vorticity, with velocity changes of 1 km s$^{-1}$
over the filament width of 1000 AU; this amounts to an impressively high
velocity gradient of 200 km s$^{-1}$ pc$^{-1}$! These large gradients
are seen only in the filaments, so we conclude that the vorticity and
the presence of thin filaments are related. As we mentioned above, the
filaments are aligned with the local magnetic field, and this
association might be an important part of the dynamics.  

Hily-Brandt suggests that the dissipation of these large gradients can
be a source of local heating that is unrelated to the usual mechanism of
photons or cosmic rays, and in particular can generate large local
temperatures that can greatly accelerate chemical processes that require
high activation energy. Observations of such chemistry come from
higher resolution observations of certain difficult-to-form molecules.

Falgarone (SINS) described the molecular maps made with the Plateau de
Beurre mm-wave array. These show HCO$^+$ with abundance far above those
predicted by equilibrium models, by factors of 10 to 100.  The lines are
much wider (several km s$^{-1}$) than the thermal width, and this is
consistent with the large velocity gradients seen in the filaments of
Hily-Brandt. Overabundances of H$_2$O and OH by similar factors are seen
by SWAS.  The chemistry of HCO$^+$ requires high abundances of CH$^+$
and $^{13}$CH$^+$, whose formation requires an activation energy of 4640
K; production of the overabundant OH and H$_2$O from O and H$_2$
requires activation energy 2980 K.

How can such temperatures be produced in a molecular cloud?  This pair
of papers by Hily-Brandt (SINS) and Falgarone (SINS) makes a strong case
that this energy comes from dissipation of the large velocity gradients
that reside in the vorticity associated with the filaments. Studies like
this are fascinating because they reveal unusual aspects of turbulence:
vorticity, large velocity gradients, dissipation, and intermittency. Do
such processes also play roles in TSAS and TSIS?

\section{ Tiny-Scale Ionized Structure} \label{tsis}

\subsection{  Classical (Tiny) Turbulence versus TSIS}

Tiny Ionized structure comes in two flavors: classical turbulence at
tiny scales (``spooks'' in this meeting's parlance) and actual
high-density structures (``things''). And also, probably, there are 
mixtures. Here we separate observed phenomena into these two categories,
but bear in mind that the separation might not be so well-defined; an
example is arclets. We should keep in mind Deshpande's (SINS) warning
that one person's structures might be another person's statistical
fluctuations

Brisken (SINS), Stinebring (SINS), and Rickett (SINS) describe the
observations and theory of arcs and substructure.  Stinebring
concentrates on observations, the other two on theory---Brisken on
details of scattering theory and Rickett more with the global
interpretation.

\subsubsection{Classical Turbulence at Tiny Scales}

Cordes (SINS) and Rickett (SINS) describe classical
scintillation. Pulsar radiation is greatly affected by its journey
through ionized gas. The simplest effects are the classical ones of
dispersion and Faraday rotation. Scintillation/scattering is more
complicated because it depends on the characteristics of electron
density fluctuations. The observable effects include angular broadening,
time variability, and frequency structure---all a result of ``spooks''.
The turbulence itself is described by the power spectral index, together
with the inner and outer scales. These scales range from a few hundred
{\it kilometers} (!) to a few hundred AU---truly tiny stuff! Moreover,
the fluctuations $\delta n_e$ aren't large, so there is no pressure
problem as we have for TSAS.

However, there were important clues in the scintillation literature,
stretching back more than 25 years, that classical turbulence is not the
whole story.  There were examples of ``tilted scintles" and ``fringing
events" that seemed to require excess bending power in the medium that
could not be provided by a Kolmogorov spectrum of density variations.
These puzzles started to clarify with the discovery of ``scintillation
arcs'' (Stinebring et al.\ 2001).  These clues form
part of the non-classical scintillation picture, which, in turn, may be
related to other long-standing puzzles, as we discuss below. They lead
to the necessity for actual {\it Structures}\dots

\subsubsection{Tiny-Scale Ionized Structures (TSIS)}

Some observational phenomena demand the presence of tiny ($\sim$~AU)
{\it structures} in the ionized gas, and we reserve the acronym TSIS
{\rm (short for TSI {\it structures})} for this stuff.  The
observational phenomona include a specific pattern of quasar flux
variability known as Extreme Scattering Events (ESEs), and also rapid
time variability known as Intra-Hour and Intra-Day Variables (IHV,
IDV). For pulsars, we see arcs and arclets; the arcs arise from ``spooks'',
and the arclets might require ``things''. We think of these effects being
produced by discrete ionized structures with large densities---and, in
contrast to the classical pulsar scintillation, they {\it do} produce a
pressure problem in the TSAS sense (Rickett, SINS).

\subsection{Observational Manifestations of Classical 
Turbulence at Tiny Scales}

\subsubsection{General Aspects; Galactic Structure}

The effect of turbulence on the scattering and scintillation properties
is described by the Fluctuation Parameter $F \propto \left({\delta n_e
\over n_e}\right)^2 / \left( f l_{outer-scale}^{2/3} \right)$ ($f$ is
filling factor), which comes from the theory of wave propagation in the
presence of fluctuations. While we (or at least some of us) can do the
nontrivial propagation theory, what we {\it cannot} do is to predict how
$F$ depends on {\it physical conditions} in the interstellar medium. In
other words, we have yet to understand how to relate physical
processes---energy input mechanisms, ionization mechanisms, shocks---to
the electron density fluctuations, their spectrum, and the inner/outer
scales. Nevertheless, the fluctuation parameter $F$ tells us a lot: it's
much bigger (factor $\sim 100$) in the inner Galaxy than in the outer,
which tells us that energy input from young stars, both living and
dying, increases turbulence. Cordes (SINS) reviews the pulsar
scintillation literature with an eye towards its dependence on Galactic
structure.

Although classical pulsar scintillation is a valuble probe of the
ionized medium, there is nothing here that requires over-pressured,
discrete structures.  A Kolmogorov turbulence spectrum (or similar)
distributed uniformly, or with some intermittency, along the line of
sight can explain most of the classical ``scintle" structure in pulsar
dynamic spectra.

We want to clarify the concept of the ``thin screen'' of scintillation
theory and observations. This terminology can be misleading.  The ``thin
screen" approximation is often invoked in scintillation theory because
it is tractable and gives some physical insight.  In the evidence above
(Stinebring, SINS) the thinness of scintillation arcs indicate that much
of the scattering is localized in ``thin screens."  These may sound like
thin, sheetlike structures whose dimension along the line of sight is
much smaller than that on the plane of the sky. This is the {\it wrong
picture!} The term ``thin screen'' means only that the ionized,
scattering medium occupies a small fraction of the line of sight. Its
actual line-of-sight length might be measured in parsecs. On the
contrary, though, the plane-of-the-sky distance that affects the
scintillation is only a few tens of AU: the scattering angles are very
small.  In other words, the bundle of scattered radiation from the
pulsar is extraordinarily thin and long and punches through any
structures along the line of sight.  So the ``thin screen'' could be
spherical, or even a cylinder whose length along the line of sight is
much, much larger than its diameter. Thus, in the context of classical
pulsar scattering observations, the term does not refer to the shape of
the scattering region, but only to the fraction of the line of sight
that it occupies.

\subsubsection{Arcs}

Dynamic pulsar spectra usually have a characteristic pattern that looks
like intersecting ocean wave trains, so it's hard to resist taking the
Fourier transform.  The 2d power spectrum of the dynamic (or primary)
spectrum is called the secondary spectrum.  It has a conjugate-frequency
axis that is proportional to differential time delay between pairs of
rays from the image (point-spread-function) and a conjugate-time axis
that is proportional to differential Doppler shift between the same
pairs of rays.  In high sensitivity observations there is almost always
faint power extending away from the origin of the secondary spectrum
along a symmetric parabolic locus (classical scintle power is localized
near the origin of the secondary spectrum plane).

A parabola arises because the differential time delay axis is
proportional to offset angle squared whereas the differential Doppler
shift axis is proportional to offset angle.  The center of the scattered
beam acts like a holographic reference signal for interference between
it and peripheral rays in a weak halo of scattered radio light.
Interference between the central part of the beam and a halo
generically produces a parabolic pattern in the secondary spectrum
because of this linear-quadratic relationship on angle (see Walker et
al.\ 2004 or Cordes et al.\ 2006 for further details).  The curvature of
the parabola depends in a simple way on location of the scattering
material along the line of sight, observing frequency, distance to the
pulsar, and pulsar proper motion.

The arc curvature can be used in cases where good estimates of the
pulsar distance and proper motion exist to determine the location of
the scattering material along the line of sight.  This determination and
the remarkable thinness of some (primary) arcs provide strong evidence that
the dominant scattering material occupies a small fraction ($\lesssim$~3~\%)
along the line of sight.  In six of the 25 or so pulsars in which scintillation
arcs have been detected, multiple arcs are seen, implying
scattering screens at different distances.  In fact, recent studies (Stinebring,
SINS) indicate that multiple arcs may be the norm and that we are
simply limited by signal strength in detecting multiple arcs along
many lines of sight.  The super sensitivity of the SKA would let us see
many more such multiple arcs and get enough statistics to solve that problem!

The parabolic arcs do not require high densities or anything exotic.
They are a generic part of small angle forward scattering in the case that
the point-spread-function (PSF) can be described in terms of a central (point-like)
beam and a wider, low-level halo.  This is the PSF produced by scattering
in a medium with Kolmogorov turbulence or any other inhomogeneity spectrum
that is shallow and produces significant wings to the image. So the
arcs are, in this meeting's parlance, ``spooks.'' 

\subsection{Observational Manifestations of Tiny Ionized Discrete
  Structures (TSIS)} \label{tsisobs}

\subsubsection{Arclets}

While scintillation arcs can be explained with Kolmogorov turbulence,
there are frequent instances when inverted parabolas or ``arclets" show
up as substructure along the main parabola.  These features appear to
require compact, relatively dense structures with a size of $\sim$~1~AU
and an electron density of $\sim$~100~cm$^{-3}$ (Hill et al.\ 2005).  In
one case, up to four discrete structures moved systematically along the
parabola---at the proper motion speed of the pulsar---while the pulsar
moved across the sky.  These definitely incur a pressure problem: it's
almost impossible to have ionization with $T \lesssim 10^4$ K, so these
must have $\widetilde P \sim 10^6$ cm$^{-3}$ K, i.e.\ they are
overpressured with respect to other thermal pressures by a factor $\sim
100$. The physical conditions required are extreme, but less extreme
than those required by Extreme Scattering Events (ESE).

In one of the highlights of the meeting, especially for the ionized
medium crowd, Brisken et al (SINS) showed a remarkable secondary
spectrum from a pulsar observed as part of a large-telescope VLBI
observation.  By using a software correlator on the baseband data they
were able to construct a filter bank with 250~Hz resolution.  The pulsar
obliged in producing arclet structure out to delays of more than 1~ms, a
world record setter!  This may not sound like a lot, but, for the known
characteristics of the scattering material along the line of sight to
this pulsar (B0834+06), this delay corresponds to an offset from the
direct line of sight of 25~mas.  Compare this with the angular radius of
the ``core" of the scattering disk of approximately 1~mas at their
observing frequency of 327~MHz!

Brisken et al.'s results showed dozens if not hundreds of extremely thin
arclets in the secondary spectrum.  Surely not all of these are discrete,
over-pressurized features in the ISM!  So, some rethinking of that 
interpretation is needed.  It looks like Brisken's ultra-thin arclets
clump into bands that, at the lower resolution reported by Hill et al.\ (2005),
may correspond to discrete scattering structures.  It would be {\it these}
that are identified with TSIS, although more work is needed to pin
this down.

\subsubsection{Extreme Scattering Events---ESEs} ESEs produce large,
rapid changes in point-source flux (Walker, SINS).  They are produced by
interstellar ionized structures that must be moving fast ($\gtrsim 100$
km s$^{-1}$), have large electron columns ($N_e \sim 10^{16}$ cm$^{-2}$)
and small sizes (a few AU), leading to large volume densities ($n_e \sim
1000$ cm$^{-3}$). Large ionization requires $T \sim 10^4$ so we have a
{\it big thermal pressure problem.}  In a new model presented at the 
meeting, Walker invoked ram pressure and
 self-gravitating molecular clouds with ionized
sheaths, whizzing through interstellar space at velocities up to 350 km
s$^{-1}$. One huge problem: to explain the observed rate, these objects
need to be {\it plentiful}: they are so small that for them to ever
produce the observed rate of Events there must be a huge number
scattered throughout interstellar space, of order $\sim 10^4$ pc$^{-3}$!
If these are truly self-gravitating molecular clouds, they contribute a
significant amount to the total interstellar mass. Walker's molecular
clouds sound like extreme versions of Heithausen's tiny molecular clouds
(SAMS), except that the SAMS don't whiz---they have small velocities.

Stinebring (SINS) emphasized that scintillation arc studies may
revolutionize our understanding of ESEs if, indeed, they turn out
to be caused by the same underlying structures in the ionized
medium.  The reason is simple.  Unlike distant quasars, pulsars
have high proper motions and scan the ISM relatively rapidly.
In addition, the information in the secondary spectrum is
interferometric in character and is influenced by scattering
structures off the direct line of sight to the pulsar.  In the
best cases, the field of view that influences scintillation arc
substructure is $\sim$~100--1000 times the angular area that
causes quasar light curve variations, and this area moves
across the sky at a rate that is typically 10 times greater.
For example, in the case of B0834+06 observed by
Brisken, the 25:1 ratio of scattering disk size to core of
the image translates into a 625:1 ratio of detection
probability since only the core of the image would undergo
an ESE if it were to intersect an over-pressurized TSIS.  

\subsubsection{ Intra-Day and Intra-Hour Variables (IDVs and IHVs)}

Lovell (SINS) described the Micro-Arcsecond Scintillation-Induced
Variability ({\it MASIV}) survey. This is an unbiased survey of 482
radio sources selected for suitability (from an original sample of
710). They used the VLA with four epochs, now going on five, and found
56\% of the sources to be variable on intra-day time scales---the {\it
Intra-Day Variables}, or {\it IDVs}.  With increasing rate of incidence,
the variability gets weaker and slower.  The variability
exhibits an annual cycle and a perfect crosscorrelation at different
observing sites with a time delay; these confirm interstellar
scintillation as the variability mechanism and their analysis produces
limits on screen velocity, scale length of scintillation pattern, and
screen distance.

Intra-Hour variables {\it IHV}s are more extreme, with faster time
scales and typically large modulations ($\sim 50\%$ in one hour). It
seems that IDVs and IHVs might be the same except for the distance of
the scattering centers, with IHVs being much closer. Given their extreme
properties, it's not surprising that they are very rare---there are only
three known. Bignall (SINS) and de Bruyn \& Macquart (SINS) discuss the
properties, which are derived from the beautifully consistent seasonal
properties and crosscorrelation time delays. 

Attaining variability on
intra-hour time scales requires that the scattering screen be nearby,
$\lesssim 30$ pc---within the Local Bubble!  Two sources, B1257-326 and
J1819+3845, are well studied. Their scattering centers lie at distances
of only 10 and 2 pc, respectively!  The scintillation length scales are
tens of thousands of km---a few Earth diameters! The transverse velocity
of the J1819+3845 scattering screen is 35 km s$^{-1}$ and its transverse
dimension lies between 60 and 6000 AU; the lower limit comes from the
number of years the variability has been observed and the upper from the
absence of variability in nearby sources. The shapes of the
scintillation patterns show that the turbulence is anisotropic by a
factor 10 or so.

\section{ Some Concluding Reflections}

\subsection{On the Electron Densities in TSIS}

Above in \S \ref{tsis}, we confidently quoted high electron densities for TSIS that
produce arclets and rapid, strong time variability. This confidence
reflects the attitude of most people at the SINS meeting, and it
certainly is justified for ESEs. However, for the other observed
phenomena some caution might be in order because estimating these
densities is not easy.

The problem is that electron density {\it fluctuations} cause the radio
wave scattering and so it is $\delta n_e / n_e$ that is being probed,
not $n_e$ alone.  Also, the analysis is complicated by lack of
information about the thickness of the scattering screen and the largest
scale of the turbulence (the outer scale) for a particular screen.
Without better information on this, it is impossible to estimate $n_e$
reliably.  About the best we can say with regard to IDVs and IHVs is
that the screens {\it may} have overpressured TSIS in them, but that it
is not {\it required} (Macquart, private communication).  If these
screens are related to the same structures causing ESEs and
scintillation arclets then there is an overpressure problem, but this is
not required in order to explain the remarkable day to hour quasar
variability.

\subsection{ On the Local Bubble}

Since kindergarten we have learned that the Local Bubble is an old
remnant of one or more explosions. It has the following basic
properties: \begin{enumerate}

\item It is full of hot gas that emits soft X-rays; in fact, this how it
was originally discovered.

\item It contains some fluff (Frisch, SINS; Linsky, SINS) in the form of
  a ``Cluster of Local Interstellar Clouds'' (CLIC), which consists of
  partially-ionized Warm Neutral Medium ($T \sim $ a few thousand K).

\item It might have enough magnetic fields to produce interesting dynamical
  effects for the local fluff (Cox \& Helenius 2003).

\item It contains material passing through the Solar System that is
  coming from the direction of the Sco/Oph association (Frisch,
  SINS). Old supernovae in this association are responsible for the
  nearby superbubble known as the North Polar Spur (Radio Loop I), so
  this approaching material is probably its edge.

\item Otherwise (being simplistic and, to some, infuriating), it is
otherwise empty and boring. \end{enumerate}

\noindent HOWEVER: \begin{enumerate}
\item Within a few pc it contains the most spectacular IHV-producing
  scattering screen (for the source J1819+3845). Given its location,
  this screen might be a boundary of one of Linsky's (SINS) clouds---or
  a common boundary of two of them, perhaps in collision.
\item Within a dozen pc it contains the second out of three
  IHV-producing screens (for the source B1257-326). All of the IHVs lie
  within the Local Bubble.
\item It  contains, within 40 pc  distance, the coldest  known HI cloud,
  which has aspect ratio like a few pieces of paper (Meyer, SINS).
\end{enumerate}

If we have all of these interesting structures so close to the Sun, and
if we are not the center of the Universe, then extreme objects like this
are much more common than we normally imagine! One more thing: do these
interesting structures exist {\it because} they are embedded in the hot
gas of the Local Bubble, or is the hot gas {\it incidental} to their
existence?

\subsection{ On TSAS vs TSIS}

HI folks, who study neutral gas, tend to think of TSAS, with its
length scales of tens of AU, as \tiny tiny\normalsize. Pulsar and
IHV/IDV folks, who study scintillation phenomena, consider the same TSAS as
\Large HUGE\normalsize \ because they concentrate on ionized gas, whose
density-dependent refractive index allows them to see truly tiny
structures. We think of the regimes as disparate, but this thinking might
simply reflect our myopic concentration on
our own areas of specialty: \begin{enumerate}

\item ISS reveals TSIS ionized turbulent scales from hundreds of KM to
hundreds of AU---a factor of $10^8$.  {\it Larger} ISS scales wouldn't
be easy to see, so pulsar and IHV/IDV folks wouldn't be aware of it.

\item TSAS observations reveal scales of tens of AU upwards. {\it
Smaller} scales wouldn't be easy to see because the observations tend to
be sensitivity limited.

\item So the observable regimes of TSAS and TSIS don't overlap. In their
  observable regimes, the TSAS and TSIS power spectra have comparable
  slopes. Are the power spectral densities---the factors out in
  front---comparable? Do TSIS and TSAS fit onto the {\it same} power
  spectrum? In other words, does the beautiful power-law spectrum for
  ionized gas of Armstrong et al.\ (1995), which covers
  a range of almost 100 km to 100 pc, apply to neutral gas as well?
\end{enumerate}

\subsection{ On Globular Clusters}

At this meeting we heard of three uses of {\it globular clusters} for
studying small scale structure in the ISM: \begin{enumerate}

\item As discussed by Ransom (SINS), globular clusters contain lots of
  evolved stars, and many neutron stars have been spun up because space
  is crowded with stars and binary systems form easily. These turn into
  pulsars, and those pulsar DMs reveal changes in $N_e$ on small angular
  scales. Most of these changes come from the foreground Galactic gas,
  not gas in the cluster, because the electron columns are not
  correlated with pulsar acceleration (and statistically, the observed
  acceleration depends on the pulsar location with respect to the cluster's
  center). 

\item Maps of optical NaI absorption lines reveal changes in $N(NaI)$ in
  the foreground gas on small angular scales. $N(NaI)$ depends on $N(HI)$
  and $n_e$ (and therefore, to some degree, $N_e$).

\item A clever new technique by Ivan King using---STARS!!---reveals
  changes in extinction, and therefore $N(HI)$, in the foreground gas on
  small angular scales.
\end{enumerate}

\noindent The relevant parties are measuring related quantities on
comparable angular scales and need to talk with one another!

\acknowledgements This research was supported in part by NSF grants
AST-0406987 (CH) and AST-0407302 (DS).


\end{document}